\documentclass[a4,10pt]{paper}

\usepackage{latexsym,enumerate}
\usepackage{amsmath,amsthm,amsopn,amstext,amscd,amsfonts,amssymb}
\usepackage{natbib,graphicx,fullpage}
\usepackage{multicol}

\title{{\bf Some discussions on the Read Paper ``Beyond subjective and objective in statistics"
by A. Gelman and C. Hennig}}

\author{{\sc Gilles Celeux$^1$, Jack Jewson$^2$, Julie Josse$^3$}\\ {\sc Jean-Michel Marin$^4$, \&~Christian P.~Robert$^{2,5}$}\\
$^1$Inria Saclay-{\^I}le-de-France,
$^2$Department of Statistics, University of Warwick\\
$^3$\'Ecole Polytechnique,
$^4$IMAG, Universit\'e de Montpellier\\
$^5$CEREMADE, Universit\'e Paris Dauphine, PSL}

\begin{document}
\maketitle
 
\begin{multicols}{2}
\section{Seconder discussion (Ch. Robert)}

While I fully agree with the authors' perspective that there are more
objectivity issues in statistics than the mere choice of a prior distribution
in Bayesian statistics, I first doubt switching terms as proposed therein will
clarify the subjective nature of the game for everyday users and second feel
there are deeper issues with the foundations of statistics that stand beyond
redemption. While surprised at seeing a paper entirely devoted to (necessarily
subjective) philosophical opinions about statistics I obviously welcome the
opportunity of such a discussion.

Indeed, ``statistics cannot do without data" but
the paper does not really broach upon the question whether or not
statistics cannot do {\em without probability}.
Although this may sound like a {\em lieu commun}, let us recall that a statistical
analysis almost invariably starts with the premise that the data is {\em random}. However,
the very notion of randomness is quite elusive, hence this aspect
truly fits within the paper topic---without even mentioning the
impossibility of establishing randomness for a given phenomenon, barring maybe
instrumental error---.  This query extends to the notion of a probabilistic
generative model and it relates more directly to the repeatability assumption that should not
be taken for granted in most realistic situations.

The central message of the paper is
that statistical analyses should be open about the many choices made in
selecting an estimate or a decision and about the forking paths of alternative
resolutions. Given that very few categories of statisticians take pride in
their subjectivity, but rather use this term as
derogatory for other categories, I fear the proposal stands little chance to
see this primacy of objectivity claims resolved, even though I agree (i) we should
move beyond a distinction that does not reflect the complexity and richness
of statistical practice and (ii) we should embrace and
promote uncertainty, diversity, and relativity
in our statistical analysis. As the discussions in Sections 2 and 5
make it clear, all statistical procedures involve subjective or
operator-dependent choices and calibration, either plainly
acknowledged or hidden under the carpet. This is why I would add (at least) two
points to the virtues of subjectivity to Section 3.3 that is
central to the paper message:
\begin{enumerate}
\item  Spelling out uncheckable assumptions about data collection;
\item  Awareness of calibration of tuning parameters.
\end{enumerate}
while I do not see consensus (item 2) as a necessary virtue. 

In fact, when going through the examination of objectivity claims by
the major threads of formalised statistical analysis, I get the feeling of
exploring many small worlds (in Lindley's words) rather than the entire spectrum 
of statistical methodologies. For instance, frequentism seems to be reduced to
asymptotics, while completely missing the area of
non-parametrics.\footnote{The later should not be considered to be ``more"
objective, but it offers the advantage of loosening model specification.} 
Frequentist inference is mostly equated with the
error-statistical proposal of Mayo (1996), despite the availability of
other and more mainstream perspectives. In particular, except for the reference to Davies (2014),
the $M$-open view seems to be missing from the picture, despite attempting to provide reasoning
{\em outside the box}. From a Bayesian perspective, the
discussions of subjective, objective, and falsificationist---missing empirical---Bayes do not really
add to the debate between these three branches, apart
from suggesting we should give up such value-loaded categories. 
I came to agree mostly with the subjectivist approach on the ground of
relativity, in that the outcome is always relative to a well-specified
Universe and that it can solely be measured in terms of that
reference. I further find the characterisation of the objectivist branch somehow
restrictive, by focussing solely on Jaynes' (2003) maxent
solution (which itself depends on many subjective choices). Hence, this section
is missing on the corpus of work about creating
priors with guaranteed frequentist or asymptotic properties. Furthermore, it
operates under the impression that an objective Bayes analysis should always
achieve the {\em same conclusion}, which misses the point of an automated derivation
of a reference prior construct. That many automations are feasible and
worth advocating nicely fits with the above relativity principle.  I
also find the defence of the falsificationist perspective, i.e. of essentially
Gelman and Shalizi (2013) both much less critical and extensive, in that,
again, this is not what one could call a standard approach to statistics.

In conclusion, the paper is appealing in calling for an end to the ``objectivier
than thou" argument, but harder to perceive as launching a move towards a
change in statistical practice. On the positive side, it exposes the need to
spell out the inputs---from an operator---leading to a statistical
analysis, both for replicability or reproducibility reasons and for
``objectivity" purposes, although solely conscious perceived choices can be
uncovered this way. It also reinforces the call for model awareness, by which I
mean a critical stance on {\em all} modelling inputs, including priors, that is, a
disbelief that any model is true, applying to statistical procedures Popper's
critical rationalism. This has major consequences on Bayesian modelling in
that, as advocated in Gelman and Shalizi (2013), and Evans (2015),
sampling and prior models should be given the opportunity to be updated when
inappropriate for the data at hand.  A potential if unrealistic outcome
of this paper would be to impose not only the production of all conscious
choices made in the construction process, but also through the posting of (true
or pseudo-) data and of relevant code for all publications involving a
statistical analysis.  On the negative side, the proposal is far too
idealistic in that most users (and most makers) of statistics cannot or would
not spell out their assumptions and choices, being unaware of or unapologetic
about these. This can be seen as a central difficulty with statistics as a
service discipline, namely that almost anyone anywhere can produce an estimate
or a $p$-value without ever being proven wrong. It is therefore hard to
fathom how the epistemological argument therein---that objective versus
subjective is a meaningless opposition---could profit statistical
methodology, even assuming the list of Section 2.3 be made compulsory.
The eight sins listed in the final section would require statistics expert
reviewers for all publications, while it is almost never the
case that journals outside our field call for statistics experts within
referees. Apart from banning all statistics arguments from journals,
I am afraid there is no hope for a major improvement in that corner.

It is thus my great pleasure to second the vote of thanks for this multi-faceted paper
that helps strengthening the foundations of our field.

\section{About Bayesian transparency (G. Celeux)}

I congratulate Andrew and Christian for their much interesting and stimulating
article. I agree with their proposition to bring to the fore the attribute {\em
transparency} instead of the attribute {\em objectivity}. \\
As a matter of fact, statistical models are not expected to explain or describe
the world, but they can rather be expected to provide tools to act on it.  For
this very reason, transparency is desirable. \\

But, I am not sure that transparency is easy to be ensured in the Bayesian
framework with complex models. Actually, the influence of hyperparameters could
be quite difficult to be analysed in an informative setting and this task could
appear to be even more difficult in a non-informative setting. In other words,
in some circumstances, choosing prior distributions could appear to be a
sorcerer's apprentice game hardly compatible with transparency. Anyhow,
transparency of a Bayesian statistical analysis requires in-depth (and
expensive) sensitivity analyses as soon as the statistical models are somewhat
complex. 

\section{Subjective Bayesian updating (J. Jewson)}

I thoroughly enjoyed how this paper brings to light the subjectivity disguised
as objectivity in statistical practise, and I relish the prospect that
understanding the impossibility of objectivity will allow researchers greater
freedom to experiment with their analysis.

Focusing on the Bayesian standpoint, I feel there is one major omission from
the author’s discussion, the methods for parameter updating. It is recognised
throughout the paper that the model used in any statistical analysis is almost
unavoidably taken to be an approximation of the decision maker's true beliefs
(or of the true data generating process depending on your perspective) (Bernardo and Smith, 2001). 
This results in statistics taking place in the $M$-complete or $M$-open world
The authors regard Bayesian updating to be
objective and transparent, suggesting that if a researcher is able to interpret
their prior, then they will by implication, be able to interpret their
posterior inference. In the $M$-closed world, I can believe this is the case.
However, in the $M$-open world Bayesian updating is less transparent. It is a
known result that Bayesian updating learns about the parameters of the
approximate model that minimise the Kullback-Leibler divergence to the data
generating process. Nonetheless, in practical terms I do not believe many
statisticians understand what it means for two distributions to be close in
terms of KL-divergence. The general Bayesian update (Bissiri et al., 2016)
reinterprets Bayes' rule as producing the Bayesian posterior attempting to
minimise the logarithmic score (and as a consequence the KL-divergence to the
data generating process):
\end{multicols}
\begin{equation}
\pi(\theta|\mathbf{x})\propto\exp(-\sum_{i=1}^n-\ell(\theta,x_i))\pi(\theta)=\exp(-\sum_{i=1}^n-\log(f(x_i;\theta))\pi(\theta)=\pi(\theta)\prod_{i=1}^nf(x_i;\theta).
\end{equation}

\begin{multicols}{2}
This provides greater transparency to the Bayesian updating process,
demonstrating that in combination with the prior, greater posterior mass is
given to parameter values whose predictions via the model $f(\cdot;\theta)$,
achieve a low logarithmic-score on the observed data $\mathbf{x}$.
Bernardo and Smith (2001) observe that scoring predictions based on the
logarithmic scoring rule places great importance on correctly specifying the
tails of the data generating process, as a large loss is incurred when an
observation predicted with low probability is seen. Bernardo and Smith (2001)
argue that tail specification is important for pure inference, but in applied
problems the statistician may require their predictions to be accurate in some
other region of the predictive distribution. The authors acknowledge that
information concerning ``how the results of an analysis are to be used or
interpreted" should form an important, subjective part of the analysis. If the
tails of the predictive distribution are important for the analysis, then the
logarithmic-score should be chosen, and this decision should be documented.
However, if the tail specification is not important, then blindly (implicitly)
using the logarithmic-score under an approximate model, can produce predictive
distributions that perform very poorly on the rest of the distribution. In this
scenario the general Bayesian update provides the tools to produce a predictive
distribution targeting an alternative loss function. 

It is tempting to try and implement the general Bayesian update without using a
model. I agree with the sentiments of the authors that using a model is
important, it provides another tool to incorporate prior information into the
analysis, and provides transparency in the way predictions are produced.
Divergence functions and their associated scores, can therefore be used to
produce model based loss functions allowing Bayesian updating to target aspects
of the posterior predictive distribution away from the tails.

In agreement with the author's recommendations concerning priors and
tuning parameters, I advocate that Bayesian updating cease to be considered
an objective black box and the room to impose subjectively is exploited and
documented.

\section{About coding (J. Josse)}

Although I agree with the authors that virtuous statistical practice involves
justifying choices made during the analysis, I do not think that statisticians
do not do it because it is subjective, but rather because no one cares enough. 
Even if such explanations are crucial, they are not valued by the community. It
is not common to have an entire article on the topic of scaling  (see. Bro \&
Smilde, 2003), and such articles are likely to be published in applied journals
not perceived to be prestigious by other colleagues. The pressure to be
published should be mentioned.  

I do not think neither that there are endless discussions on the subject of
objectivity, subjectivity, but the fact that there are many ways to deal with a
problem will always lead to this impression of subjectivity. This debate seems
more linked to the Bayesian literature, perhaps because it has at least the
merit of questioning what information is incorporated into the analysis. This
could explain why the ones who use it for mathematical simplicity, which is
quite justifiable, may be seen as "opportunistic Bayesians". It is crucial to
make choices clear.

The choice of data coding is important. In sensory analysis, there is debate as
to whether the Likert scale should be coded as quantitative or qualitative. To
be ``coding free", some methods  (Pages, 2015) consider a compromise
between these two points of view and  highlight the specificity of each. 
The example of clustering is striking. Callahan  \textit{et al.} (2016) also
stresses the need to document analyses with a view to reproducibility. He has
shown that there could be "more than 200 million possible ways of analysing
these data". Of course, there is no ``good"  solution, it ``depends" on the
characteristics of the data one wants to capture. 

Even when a problem is well-characterised, two statisticians who make use of
the same data will use different approaches. This is mainly due to their
personal history \footnote{On a personal basis, I speak about "a statistical
sensitivity".}, and the expression "when one has a hammer, one sees nails
everywhere" often applies. This is not necessarily a problem, and experience
gained must be used. As mentioned by the authors, collaboration should be
encouraged, for example in the development of simulation studies.

\begin{quote}{\em 
``The best future is one of variety not uniformity".} John Chambers
\end{quote}

In conclusion, this paper has the merit of promoting transparency, awareness of
the limits of a study, and its context-dependent nature \footnote{As a French
statistician, I also appreciated that the paper begins with with "we cannot do
statistics without data".}. The battle is not lost because the community is
already encouraging the sharing of code and data. It is worth remembering that
different points of view can be legitimate.  

\section{About randomness (J.-M. Marin, J. Josse and C. Robert)}

We congratulate the authors on their exposition of the issues of modelling and experimenters' input on statistical inference and welcome this opportunity to discuss some fundamentals on such neglected topics.

A first criticism is about the focus that is definitely set on (statistical) models and
the ensuing (statistical) inference. We indeed wonder if this focus is {\em de
facto} set on a completely inappropriate problematic, namely arguing between
ourselves [meaning academic statisticians] about the best way to solve the
{\em wrong} problems, while the overwhelming majority of users is more than ready 
to buy and exploit quick-and-dirty solutions, provided these carry a sufficient modicum of
efficiency, i.e., ready to enforce imprecise and suboptimal inference. Taking,
for instance, the perspective of an Internet ordering operator seems much more relevant
than focussing on the statistical background for validating the existence of an
elementary particle. In other words, there are many more immediate
(production) problems that call or even scream for statistical processing than
well-set scientific questions.

We adhere to the argument that the scientific realism position allows for a more workable {\em
modus operandi}.) This particularly applies to data
analyses in social sciences and medicine, as opposed to hard sciences where
(almost) all experimental conditions can be expected to stay under control or at
least to be stationary across repeated experiments. Maybe not-so-incidentally, the
three examples treated in Section 4 belong to the former category.\footnote{We
find it highly significant and rather amusing that the authors picked a
clustering example as clustering is certainly one of the statistical procedures
that is the most loaded with preconceptions and forking interpretations.}
These examples are all worth considering as they bring more details,
albeit in specific contexts, on the authors' arguments. However, most of them
give the impression that the major issue in the debate does not truly stands 
with the statistical model itself, referring instead to a concept of model that
only is relevant for hard sciences. This was further illustrated by the mention
of outliers during the talk, a notion that is nonsensical in an $M$-open
perspective. It is obviously illusory to imagine the {\em all models are wrong}
debate settled but it would have been relevant to see it directly addressed.

As a side remark, the rather hasty dismissal of machine learning in Section 5
is disappointing, because there is at least one feature for which machine
learning tools are worth considering, and it is that they avoid leaning too
much on a background model, using instead predictive performances as an
assessment criterion. The alluring almost universal availability of such tools,
as well as the appearance of objectivity produced by the learning analogy,
could have been addressed in the spirit of the paper, especially in a context
of those techniques taking over more traditional statistical learning in many
areas.

Finally,  the powerful role of software should  be mentioned. Indeed,  the
availability of methodology in software may explain why certain practices, even
when flawed, are still in use.  Even though it is difficult to imagine
software without any default values for "tuning" parameters, solutions may be
envisaged to force users to be aware of the underlying choices and to assume
them.  In addition, all-inclusive statistical solutions, used through
``point-and-shoot" software by innumerate practitioners in mostly inappropriate
settings, give them the impression of conducting ``the" statistical analysis.
This false feeling of ``the" proper statistical analysis and its relevance for
this debate also transpires through the treatment of statistical expertises by
media and courts, as well as some scientific journals.

\section*{References}

\begin{itemize}
\item[]{\sc Bissiri, P., Holmes, C., and Walker, S. G.} (2016) A general framework for
updating belief distributions. \textit{Journal of the Royal Statistical Society:
Series B}.
\item[]{\sc Bernardo, J.M., and Smith, A.F.} (2001). {\em Bayesian Theory}. IOP Publishing.
\item[]{\sc Bro, R. and Smilde, A.K.} (2003). Centering and scaling in component analysis, \textit{Journal of Chemometrics.}  
\item[]{\sc Callahan, B., Proctor, D., Relman, D., Fukuyama, J., and Holmes, S.} (2016). Reproducible research workflow in R for the analysis of personalized human microbiome data. \textit{Pac Symp Biocomput.}
\item[]{\sc Jaynes, E.} (2003). {\em Probability Theory}. Cambridge University Press, Cambridge.
\item[]{\sc Mayo, D. G.} (1996). {\em Error and the Growth of Experimental Knowledge}. University of Chicago Press.
\item[]{\sc Pages, J.} (2015) \textit{Multiple Factor Analysis by Example Using R}, Chapman \& al.
\end{itemize}

\end{multicols}
\end{document}